\documentclass [a4paper,12pt]{article}
\usepackage{amssymb}
\usepackage{graphicx}
\usepackage{amsmath}
\usepackage{fancybox}
\usepackage{color}
\usepackage{geometry}

\newcommand{\be}{\begin{equation}}
\newcommand{\ee}{\end{equation}}
\newcommand{\bea}{\begin{eqnarray}}
\newcommand{\eea}{\end{eqnarray}} 
\begin{document}
\setlength{\baselineskip}{18pt}
\begin{titlepage}

\begin{flushright}
KOBE-TH-19-06       
\end{flushright}
\vspace{1.0cm}
\begin{center}
{\LARGE\bf 't Hooft-Polyakov monopole and instanton-like topological solution in gauge-Higgs unification} 
\end{center}
\vspace{25mm}

\begin{center}
{\large
Kouhei Hasegawa 
and 
Chong-Sa Lim$^*$
}
\end{center}
\vspace{1cm}
\centerline{{\it
Department of Physics, Kobe University,
Kobe 657-8501, Japan.}}

\centerline{{\it
$^*$
Department of Mathematics, Tokyo Woman's Christian University, Tokyo 167-8585, Japan }}
%
%

\vspace{2cm}
\centerline{\large\bf Abstract}
\vspace{0.5cm}

We consider 't Hooft-Polyakov monopole (TPM) as a topological soliton in the 5-dimensional (5D) theory of gauge-Higgs unification. This scenario provides a very natural framework to incorporate the TPM, since the adjoint scalar is builtin as the extra-space component of higher dimensional gauge field. In the process of the analysis, we realize that the condition to be satisfied by the Bogomolny-Prasad-Sommerfield (BPS) state of TPM, the BPS monopole, is equivalent to an (anti-)self-dual condition for the higher dimensional gauge field. This observation, in turn, suggests the presence of instanton-like topological soliton living in the 4D space (including the extra dimension), instead of the 4D space-time in the case of ordinary instanton, say ``space-like instanton" with finite energy, instead of finite action. We construct the field configuration for the space-like instanton and calculate its mass, handled by the compactification scale. Next we discuss the BPS monopole, as an anti-self-dual gauge field. We start by constructing a hedgehog-type solution as what is obtained by a local gauge transformation from a trivial vacuum. 
We also argue in some detail that the relation between these two types of topological solitons becomes manifest through the unified description by use of the ansatz adopted by 't Hooft.

\end{titlepage}

\section{Introduction} 

The standard model has two theoretical problems in its Higgs sector: \\ 
(1) The gauge hierarchy problem.  \\ 
(2) The strengths of the Higgs boson interactions, such as Yukawa coupling constants, cannot be predicted theoretically. \\ 
The first problem gave the main motivation to search for physics beyond the standard model (BSM). 

The essential problem hidden behind these two is that there is no guiding principle to restrict the Higgs interactions. 
Thus, it would be nice if Higgs interactions can be controlled by gauge principle. This is the philosophy of gauge-Higgs unification (GHU), a possible scenario of BSM, where the origin of Higgs boson is gauge boson. To be more precise, the Kaluza-Klein (KK) zero mode of the extra space component of higher dimensional gauge field is identified with the Higgs field \cite{Manton:1979kb}, \cite{Hosotani}. Concerning the problem (1) mentioned above, by virtue of higher dimensional local gauge symmetry, the hierarchy problem is solved \cite{Hatanaka:1998yp}, thus opening a new possibility to construct 
realistic models of particle physics based on this scenario.  

One may wonder whether such identified Higgs field is physically meaningful. In this scenario the vacuum expectation value (VEV)
of the Higgs field is nothing but a constant gauge field, giving vanishing field strength, and therefore seems to be physically meaningless. Fortunately the answer to the raised question is yes. Let us recall that there is a remarkable case where a constant gauge field plays an important physical role, even if the field strength vanishes in the space of our interest, i.e. the Aharonov-Bohm (AB) effect. In GHU, the Higgs field may be regarded as a sort of AB (or Wilson-line) phase, which in turn means that all physical observables are periodic in the Higgs field, thus leading to ``anomalous" Yuakawa couplings \cite{Hosotani:2008tx}, \cite{Hasegawa:2012sy}, for instance. As is well-known, the key ingredient of the AB effect is that the space is non-simply connected. Similarly, in 5-dimensional (5D) GHU model with a circle as its extra space, the fact that the circle is a non-simply connected space makes the Higgs field physically meaningful. The lesson here is that in GHU the magnetic property of gauge field, together with the topological nature of the extra space, is quite essential for the scenario to work.  

Concerning the problem (2) mentioned above, it has been pointed out that GHU also has an interesting implication for Yukawa couplings and therefore for fermion masses. Namely, it has been demonstrated in the framework of 6D GHU with a torus as its extra space that the observed impressive hierarchical structure of fermion masses can be naturally understood without fine tuning of the parameter by the topological nature of field configuration generated by magnetic monopole placed inside the torus \cite{Lim}. In this mechanism, the quantization condition of magnetic charge $g_{m}$, $gg_{m} = \nu$ ($\nu$: integer, \ $g$: gauge coupling constant), plays a crucial role, which is the reflection of the topological nature of gauge field configuration. In fact, the non-trivial topological configuration of the gauge field is characterized by homotopy class (group) $\pi_{1}(U(1)) = {\bf Z}$ (for Dirac type magnetic monopole), with ${\bf Z}$ standing for the integer $\nu$ in the quantization condition mentioned above. The index theorem by Atiyah-Singer applied for the 2D torus as the extra space is also helpful to understand the mechanism to explain the hierarchical fermion masses:       
\be 
\label{1.1}
{\rm The \ index \ of \ Dirac \ operator} = \frac{1}{2\pi} \int_{{\rm torus}} {\rm tr} \ F,  
\ee 
where the right-hand side indicates total magnetic flux penetrating the torus and this equation is another proof of the quantization of magnetic charge, as the left-hand side should be an integer, say $\nu$. This also guarantees the appearing of $\nu$ chiral fermions, $\nu$ ``generations" of fermion.  

In \cite{Lim}, the presence of Dirac-type magnetic monopole was just assumed and its origin was not specified.  It would be desirable if the Dirac-type monopole can be replaced by the 't Hooft-Polyakov monopole (TPM) \cite{'t Hooft}, \cite{Polyakov}, which does not cause any singularity in the gauge and scalar field configurations.   
   
It should be emphasized that GHU actually provides us with a very natural framework to incorporate the TPM:  
the scalar field belonging to the adjoint representation of the gauge group, inevitable to realize TPM, necessarily exists in GHU as the extra space component of the gauge field, $A_{y}$. It is also worth noticing that in the minimal electro-weak unified model based on the GHU scenario, the SU(3) GHU model \cite{Kubo}, the simple gauge group inevitably leads to the emergence of the TPM, just as in the case of SU(5) grand unified theory. 

Thus, the original purpose of this paper is to formulate the TPM in the framework of GHU, and see whether there appear some genuine features, characteristic to GHU. The model we adopt is 5D SU(2) GHU model. Now the field configuration of TPM is obtained as a topologically non-trivial classical solution in the 5D pure Y-M theory, without any necessity to introduce matter fields. The TPM is builtin in this model.

Interestingly, in the process of the analysis we have found, as demonstrated below, that the condition to be satisfied by the Bogomolny-Prasad-Sommerfield (BPS) state of the TPM, say ``BPS monopole", is exactly equivalent to the (anti-) self-dual condition for the space-like components of higher dimensional field strength, subject to the naive dimensional reduction, in GHU. This observation, in turn, strongly suggests the presence of instanton-like topological soliton living in the 4D space (including the extra dimension), instead of the 4D space-time in the case of ordinary instanton, say ``space-like instanton" with finite energy, instead of finite action. We will construct the gauge field configuration of the space-like instanton.  

At low energies, we expect that only KK zero modes are relevant, assuming all fields are independent of the extra space coordinate $y$: the naive dimensional reduction. So, the BPS monopole solution, independent of $y$, with the isotropy of 3D space is expected to be the low energy version of the space-like instanton solution having 4D isotropy. Since the dimensional reduction is expected to be realized by a smooth modification of gauge field, the homotopy classes responsible for these two types of topological solitons are expected to be identical. In fact, the homotopy class responsible for the space-like instanton, $\pi_{3}(SU(2)) = {\bf Z}$, and the one responsible for the TPM, $\pi_{2}(SU(2)/U(1)) = {\bf Z}$, are identical.  

By the way, in the ordinary 4D space-time, the BPS monopole is a classical solution of the equation of motion (E.O.M.) only when the scalar potential is small enough. In the 5D GHU, however, the BPS monopole is the solution of the E.O.M. automatically, as is guaranteed by the Bianchi identity. Also note that the Higgs potential is not allowed at the classical level and is generated only through the quantum effects in 5D GHU. After constructing the gauge field configuration of the BPS monopole as a (anti-)self-dual field with finite energy, relying on the SO(3) symmetry of the 3D space, we will argue in some detail the relation between the space-like instanton and the BPS monopole by utilizing the ansatz adopted by 't Hooft \cite{'thooftansatz} in order to construct (anti-)self-dual fields.

\section{``Space-like instanton" in 5D space-time} 

We start from the action of the SU(2) gauge theory in ordinary 4D space-time, which leads to the TPM: 
\be 
\label{2.1} 
S =\int {\cal L} \ d^{4}x, \ \ 
{\cal L} = - \frac{1}{2}{\rm Tr}(F_{\mu \nu}F^{\mu \nu}) + 
 {\rm Tr}\{(D_{\mu}\phi)(D^{\mu}\phi) \} - V(\phi) 
\ee 
where $\phi = \phi^{a} (\frac{\tau_{a}}{2})$ is a scalar field, introduced as a matter field, belonging to the adjoint (triplet) representation of SU(2), with $\phi^{a}$ being real scalar fields and $\tau_{a}$ denoting Pauli matrices. Also,      
\bea 
&& A_{\mu} = A^{a}_{\mu} (\frac{\tau_{a}}{2}), \ \ F_{\mu \nu} = \partial_{\mu}A_{\nu} - \partial_{\nu}A_{\mu} -ig [A_{\mu}, A_{\nu}],  \\ 
&& D_{\mu} \phi = \partial_{\mu} \phi -ig [A_{\mu}, \phi], 
\label{2.2} 
\eea 
and $V(\phi)$ is the scalar potential. 

On the other hand, in the framework of 5D SU(2) GHU model, we do not have to introduce the adjoint scalar field $\phi$, since its role is played by the extra-space component of the gauge field, $A_{y}$. The 5D action corresponding to (\ref{2.1}) is just that of the 5D SU(2) pure Yang-Mills theory: 
\be 
\label{2.3} 
S =\int {\cal L} \ d^{4}x dy, \ \ {\cal L} = - \frac{1}{2}{\rm Tr}(F_{MN}F^{MN}),  
\ee 
where $M, N = 0, 1, 2, 3, y$. The extra dimension is assumed to be a circle with radius $R$, and $y$ is the coordinate along the circle. Under naive dimensional reduction, realized by just ignoring the $y$-dependence of the fields (corresponding to taking only the KK zero modes of the fields into account), the action (\ref{2.3}), after the $y$-integral and suitable rescaling of the fields and the gauge coupling by a factor $\sqrt{2\pi R}$ to adjust their mass dimensionality, just reduces into (\ref{2.1}), identifying $A_{y}$ with $\phi$, but without the potential term $V(\phi)$. This is easily seen by noting $F_{MN}F^{MN} = F_{\mu \nu}F^{\mu \nu} + 2 F_{\mu y}F^{\mu y}, \ F_{\mu y} = \partial_{\mu} A_{y} -ig [A_{\mu}, A_{y}] \ (\mu, \nu = 0, 1, 2, 3)$ under the naive dimensional reduction. 

For the TPM in 4D space-time with a given (quantized) magnetic charge 
$g_{m} = \frac{\nu}{g} \ (\nu: {\rm integer})$, finite energy classical solutions 
for the E.O.M. of fields are generally obtained only numerically. 
But, there is a limiting case where we get analytic solutions, i.e. BPS states for the TPM called BPS monopoles. It should be noted that for such BPS monopole to be the (approximate) solution of the E.O.M., the scalar potential should be small enough. This is automatically guaranteed in the 5D GHU model, where the potential for $A_y$ only arises at the quantum level and therefore can be ignored at the classical level. 

The condition to be satisfied by the BPS monopole is 
\be 
\label{2.4} 
F_{ij} = \pm \epsilon_{ijk} D_{k}\phi \ \ (i,j,k = 1,2,3),   
\ee 
with $\epsilon_{123} = 1$. Interestingly, we realize that identifying $F_{ky}$ with $D_{k}\phi$ under the naive dimensional reduction, (\ref{2.4}) is nothing but 
the self-dual or anti-self-dual condition for the spacial components of higher dimensional gauge field in the 4D space including the extra dimension, instead of the (Euclidean) 4D space-time:  
\be 
\label{2.5} 
F_{IJ} = \pm \frac{1}{2}\epsilon_{IJKL} F_{KL} \ \ (I,J,K,L = 1,2,3,y),    
\ee 
with $\epsilon_{123y} = 1$. (In the process of preparing this manuscript we have realized that a similar observation was made in the literature \cite{Manton_self-duality}, \cite{Manton}, \cite{Manton2},
but in the framework of the gauge theory in the ordinary 4D space-time, not in the framework of GHU). 

This observation, in turn, strongly suggests the presence of instanton-like topological soliton, but living in the 4D space  including the extra dimension, instead of the 4D space-time in the case of the ordinary instanton, say ``space-like instanton", whose field configuration is a topologically non-trivial (anti-)self-dual gauge field with finite energy, instead of finite action in the case of the ordinary instanton.   

The assertion given above can be explicitly confirmed as follows. First, the Hamiltonian for static field configuration is 
\be 
\label{2.6}
H = \int {\cal H} \ d^{3}x dy, \ \ \ {\cal H} = \frac{1}{2}{\rm Tr}(F_{IJ}^{2}) 
= \frac{1}{4}{\rm Tr}\{ (F_{IJ} \pm \tilde{F}_{IJ})^{2} \mp 2 F_{IJ}\tilde{F}_{IJ} \},  
\ee 
where $\tilde{F}_{IJ} \equiv \frac{1}{2} \epsilon_{IJKL} F_{KL}$. The first term in the right-hand side of (\ref{2.6}) vanishes provided the (anti-)self-dual condition 
\be 
\label{2.7}
F_{IJ} = \pm \tilde{F}_{IJ}
\ee 
is met. (\ref{2.7}) is nothing but the condition (\ref{2.5}). Also, the wisdom from the argument of the instanton tells us that the remaining second term in the right-hand side of(\ref{2.6}) is a topological invariant and is given by an integer, known as Pontryagin index:  
\be 
\label{2.8}
\nu = \frac{g^{2}}{16\pi^{2}} \int {\rm Tr}(F_{IJ}\tilde{F}_{IJ}) \ d^{3}x dy.  
\ee 
Thus, for fixed $\nu$, the field configuration of space-like instanton satisfying the (anti-)self-dual condition (\ref{2.7}), gives the minimum energy of 
\be 
\label{2.9}
M_{\nu} = \frac{8\pi^{2}}{g^{2}}|\nu| = \frac{4\pi |\nu|}{g_{4}^{2}} \frac{1}{R}, 
\ee 
where $g_{4} = g/\sqrt{2\pi R}$ is the 4D gauge coupling. $M_{\nu}$ of the order $\frac{M_{c}}{\alpha}$ ($M_{c} = 1/R$: compactification scale, \ $\alpha$: fine structure constant) should be regarded as the mass of this particle-like soliton. The (static) field configuration of the space-like instanton  (with $A_{0} = 0$) is the exact solution of the E.O.M. as is easily seen by use of the Bianchi identity.

We now discuss the field configuration of $A_M$ to describe the space-like instanton with the Pontryagin index $\nu = -1$, for instance.  Since the extra-space is assumed to be compactified on a circle S$^1$, $A_{M}$ should satisfy the periodic boundary condition, i.e. $A_{M}(y = \pi R) = A_{M}(y = - \pi R)$.  
We wish we could solve exactly the anti-self-dual condition (\ref{2.7}) for the arbitrary size $R$ of the extra space, by performing the Kaluza-Klein (KK) expansion of $A_{M}$ and solving the condition (\ref{2.7}) for each of the KK mode. Unfortunately, we have not succeeded to get the exact solution: the condition leads to coupled differential equations for KK modes, as pointed out at the last paragraph of the ``Summary and Discussion", which are not easy to solve. To our knowledge, the exact analytic solution has not been found to this date.

We also would like to point out that in the literature there have been attempts to solve the (anti-)self-dual condition at the finite temperature Yang-Mills theory, not in 5D space-time, but in ordinary 4D space-time. The (anti-)instanton at finite temperature called ``caloron" \cite{Harrington}, \cite{Dunne} has some similarity to the space-like instanton with finite $R$, as we realize by replacing the temperature $T$ by $1/R$, although in the case of finite temperature field theory an anti-periodic boundary condition is often relevant, instead of the periodic boundary condition of our interest here. Again, to our knowledge, the gauge field configuration for the caloron has not been obtained analytically, though there exists an argument based on numerical (lattice) calculation \cite{Perez}.

Under such circumstance, in this paper we consider the extreme case $R \to \infty$, namely the de-compactification limit, leaving the issue of the exact solution of the anti-self-dual condition in the realistic case of finite $R$ for future study.  In this de-compactification 
limit, the boundary condition reduces to $A_{M}(y = \infty) = A_{M} (y = -\infty)$, and the relevant homotopy group is $\pi_{3}(SU(2))$, just as in the case of ordinary 4D instanton.  Thus the solution with $\nu = -1$ is easily known to be  
\be 
\label{2.10}
gA_{I} = i(\frac{\rho^{2}}{\rho^{2} + \lambda^{2}}) U^{-1} \partial_{I} U \ \ \ (U =  i \hat{x}_{I}\bar{\sigma}_{I}),   
\ee 
where $\hat{x}_{I} = \frac{x_{I}}{\rho} \ (\rho \equiv \sqrt{x_{I}^{2}} = \sqrt{x_{i}^{2} + y^{2}})$ and $\bar{\sigma}_{I} = (\vec{\tau}, -i I)$ with $I$ being 2$\times$2 unit matrix, while the arbitrary parameter $\lambda$ denotes the spacial size of the space-like instanton. More explicitly, the ``hedgehog-type" field configurations are   
\bea 
&& g\vec{A} = - \frac{y\vec{\tau} + \vec{\tau} \times \vec{x}}{\rho^{2} + \lambda^{2}}, 
\label{2.10a} \\ 
&& gA_y = \frac{\vec{\tau}\cdot \vec{x}}{\rho^{2} + \lambda^{2}}. 
\label{2.10b}
\eea  
It is easy to check that this gauge field is anti-self-dual (with $\nu = -1$) and therefore describes one space-like anti-instanton. The solutions (\ref{2.10a}) and (\ref{2.10b}) clearly satisfy the condition $A_{M}(y = \infty) = A_{M}(y = -\infty) \ ( = 0)$, with $M = 1, 2, 3, y$.

\section{The BPS monopole}

The equivalence of the condition (\ref{2.7}) with the condition (\ref{2.4}) under the naive dimensional reduction implies that the BPS monopole in 3D space is closely related to the space-like instanton in 4D space. In fact, the homotopy class responsible for the space-like instanton, $\pi_{3}(SU(2)) = {\bf Z}$ and the one responsible for the TPM, $\pi_{2}(SU(2)/U(1)) = {\bf Z}$, are identical.   

The mutual relation, however, is not apparent. As we have already seen above, the hedgehog-type space-like instanton clearly has a combined symmetry of SO(4) rotation of 4D space and SO(4) as the isometry of SU(2) internal gauge space ($\simeq$ S$^{3} \simeq$SO(4)/SO(3)), while the BPS monopole has a combined symmetry of SO(3) rotation of 3D space and SU(2) gauge transformation, as is shown below. So, in this section, we just search for the BPS monopole solution of winding number $1$, keeping only the $y$-independent KK zero modes justified under the naive dimensional reduction, invoking the combined symmetry of SO(3) rotation of 3D space and SU(2) gauge symmetry as the guiding principle in the framework of SU(2) GHU model. Further discussion concerning the relation between the space-like instanton and the BPS monopole will be given in the next section.

\subsection{Lessons from simplified mapping}

In order to understand the behavior of hedgehog-type finite energy solution, we first consider a simplified mapping from $S^{1}$ on the $x-y$ plane, instead of $S^{2}$ in 3D space of our real interest, to the SO(2) rotation among $(A^{1}_y, A^{2}_y)$ around the third axis in the internal space, the subgroup of SO(3) ($\sim$ SU(2)), of our real interest. 
The fields and gauge coupling appearing throughout this section are supposed to be those in ordinary 4D space-time, obtained under naive dimensional reduction, though the combination 
$gA_{M}$ are invariant under the rescaling by the factor $\sqrt{2\pi R}$ and therefore we do not need to specify the space-time dimension.   

The gauge group SO(2) ($\simeq$ U(1)) is completely broken by the assumed VEV, $|\langle A_y \rangle | = v$. A point $(x, y)$ on the circle on the $x-y$ plane corresponds to $z = x + iy = r e^{i\varphi}$ ($r, \ \varphi$ : polar coordinates) on a complex plane, and accordingly $(A^{1}_y, A^{2}_y)$ is combined into a complex field $A_{y} \equiv A^{1}_{y} + i A^{2}_{y}$. The hedgehog-type configuration of the field with winding number 1 is  
\be 
\label{3.1}
A_{y}(r e^{i\varphi}) = ve^{i\varphi} \ \ {\rm or} \ \ A^{a}_{y} = v \hat{x}_{a} \ \ (a = 1, 2),  
\ee 
with $\hat{x}_{a} = \frac{x_{a}}{r}$ and $x_{1} = x, \ x_{2} = y$. 
This $\varphi$ dependent field configuration leads to a non-vanishing kinetic term and will not lead to the finite energy solution. If, however, this configuration can be regarded as what is obtained by a local gauge transformation from a trivial vacuum state with vanishing energy (assuming that for the VEV the potential $V(A_y)$ vanishes),   
\be 
\label{3.2}
A_{0y} = v, \ \ A^{3}_{0\mu} = 0,  
\ee 
we expect that the obtained field configuration leads to finite energy, even though non-vanishing for the non-trivial topological configuration.

Obviously, (\ref{3.1}) is obtained from (\ref{3.2}) by U(1) transformation with a group element $U = e^{i\varphi}$. Accordingly, this local gauge transformation yields a configuration of 3D spacial gauge field (still keeping $A_{0} = 0$), 
\be 
\label{3.3}
gA^{3}_{i} = i U^{\ast}\partial_{i}U = \frac{1}{r}\epsilon_{3ij}\hat{x}_{j},   
\ee 
which are obtained by use of $\frac{\partial \varphi}{\partial x_{1}} = - \frac{x_{2}}{r^{2}}, \ \frac{\partial \varphi}{\partial x_{2}} = \frac{x_{1}}{r^{2}}$.  (\ref{3.1}) and (\ref{3.3}) form the field configuration with winding number 1. We will see in the next subsection that what we obtain for $A^{a}_{i}$ in the realistic case of SU(2), for large $r$, is the generalization of (\ref{3.3}) (see (\ref{3.16})). 

Such obtained gauge field (\ref{3.3}) should have singularity somewhere, since otherwise it would be gauge-equivalent to (\ref{3.2}) with trivial topology (with winding number 0). In fact, (\ref{3.3}) is, say ``almost pure gauge" configuration, giving vanishing field strength almost everywhere, except for the origin, where it is divergent.

\subsection{SU(2) GHU model}   

Now we turn to the topologically non-trivial hedgehog-type finite energy solution in SU(2) GHU, of our real interest. 
First, to get finite energy solution, we first consider the asymptotic behavior of gauge fields for large $r$ ($r$: the distance from the origin). The hedgehog-type field configuration of $A_{y}$ is given by   
\be 
\label{3.4} 
A_{y} = v \hat{x}_{a} (\frac{\tau_{a}}{2}) \ \ (\hat{x}_{i} \equiv \frac{x_{i}}{r}, \ r = \sqrt{x_{i}^{2}}),   
\ee 
or 
\be 
\label{3.4a} 
A^{a}_{y} = v \hat{x}_{a},  
\ee
which is the generalization of (\ref{3.1}) with $a$ now taking all of $1, 2, 3$. The VEV $v$ of $A_{y}$ causes the spontaneous symmetry breaking (SSB), SU(2)$\to$U(1).

The field configuration of $A_{i} \ (i = 1, 2, 3)$ is obtained by imposing the condition $F_{iy} = 0$, which is necessary to get a finite energy  solution:  
\be 
\label{3.5} 
F_{iy} = \partial_{i}A_{y} - ig [A_{i}, A_{y}] = 0,  
\ee 
where we have used $\partial_{y}A_{i} = 0$, valid for the KK zero mode. (\ref{3.5}) may be written in terms of component fields $A^{a}_{i}, \ A^{a}_y$ of adjoint representation as follows,  
\be 
\label{3.6} 
\partial_{i}A^{a}_{y} + g \epsilon_{ab'c'} A^{b'}_{i} A^{c'}_y = 0.  
\ee 
Multiplying this equation by $\epsilon_{abc}$ and using $\epsilon_{abc}\epsilon_{ab'c'} = \delta_{bb'}\delta_{cc'} - \delta_{bc'}\delta_{cb'}$, we get 
\be 
\label{3.7} 
\epsilon_{abc}\partial_{i}A^{a}_{y} + g (A^{b}_{i}A^{c}_{y} - A^{c}_{i}A^{b}_{y}) = 0.  
\ee 
Then we multiply by $\frac{\hat{x}_{c}}{gv}$ and use (\ref{3.4a}) to get, by noting $\hat{x}_{c}^{2} = 1$,   
\be 
\label{3.8} 
\frac{1}{g}\epsilon_{abc} \hat{x}_{c} \partial_{i} \hat{x}_{a} + A^{b}_{i} - A^{c}_{i} \hat{x}_{c} \hat{x}_{b} = 0.  
\ee 
If we ignore the third term in the left-hand-side of (\ref{3.8}), which is justified later on, we obtain 
\be 
\label{3.9}
A^{b}_{i} = - \frac{1}{g} \epsilon_{abc} \hat{x}_{c} \partial_{i} \hat{x}_{a} = - \frac{1}{gr}\epsilon_{ibc}\hat{x}_{c}, 
\ee 
i.e. 
\be 
\label{3.10} 
gA^{a}_{i} = \frac{1}{r}\epsilon_{aij}\hat{x}_{j}.  
\ee 
Then the third term of (\ref{3.8}) turns out to vanish, since $A^{c}_{i} \hat{x}_{c} \hat{x}_{b} = \frac{1}{gr}\epsilon_{cij}\hat{x}_{j}\hat{x}_{c}\hat{x}_{b} = 0$. Thus the manipulation we took above (to ignore the third term) is known to be justified. Note that (\ref{3.10}) is the natural generalization of (\ref{3.3}) and has manifest covariance under the combined symmetry SO(3). 

The derived field configuration (\ref{3.10}) is nothing but what was proposed by 't Hooft and Polyakov \cite{'t Hooft}, \cite{Polyakov} (see also \cite{Weinberg}):   
\be 
\label{3.16} 
gA_{i} = \frac{1}{r}\epsilon_{aij}\hat{x}_{j}(\frac{\tau_{a}}{2}). 
\ee 
Eq.s (\ref{3.4}) and (\ref{3.16}) are our final results.

Once we get the behaviors of the fields for large $r$, the derivation of the BPS state for arbitrary $r$ is straightforward \cite{Weinberg}. Let $F(r), \ G(r)$ be defined as 
\bea 
&& A_{y} = F(r) v \hat{x}_{a} (\frac{\tau_{a}}{2}), \nonumber \\ 
&& gA_{i} = G(r) \frac{1}{r}\epsilon_{aij}\hat{x}_{j}(\frac{\tau_{a}}{2}),  
\label{3.18}
\eea 
with $F(\infty) = G(\infty) = 1$ as is required by (\ref{3.4}) and (\ref{3.16}). 
Then, the condition to be satisfied by the BPS monopole, (\ref{2.4}) with $\phi$ being identified with $A_y$, is equivalent to the following coupled differential equations 
\bea
&& gv F(1 - G) =  \frac{dG}{dr}, \nonumber \\  
&& evr^{2} \frac{dF}{dr} = G(2 - G). 
\label{3.18'} 
\eea 
Solving these equations under the boundary conditions $F(\infty) = G(\infty) = 1$, we get  
\bea
&& F(r) =  \coth \tilde{r} - \frac{1}{\tilde{r}}, \nonumber \\  
&& G(r) =  1 - \frac{\tilde{r}}{\sinh \tilde{r}}, 
\label{3.19} 
\eea 
with $\tilde{r} = gvr$. 

The possessed energy and therefore the mass of this BPS monopole is known to be $M_{{\rm BPS}} = \frac{4\pi}{g_{4}}v$. Let $\frac{g_{4}}{2}v$ be identified with the mass $M_{X}$ of a massive gauge boson {$X$}, which acquires the mass through the VEV $v$. For instance, in the case of the realistic SU(3) GHU model \cite{Kubo}, $M_{X}$ is the mass scale of the SSB, SU(3)$\to$SU(2)$_{L} \times $U(1)$_Y$, supposed to be much larger than the weak scale $M_{W}$. Then, $M_{{\rm BPS}} = \frac{8\pi}{g_{4}^{2}}M_{X} \sim \frac{M_{X}}{\alpha}$.

\section{The relation between the space-like instanton and the BPS monopole} 

Though both field configurations of the space-like instanton and the BPS monopole can be interpreted as (anti-)self-dual gauge field, 
to establish the mutual relation is not straightforward, as we mentioned in the previous section.   

However, a unified description of both field configurations turns out to be possible by utilizing the ansatz adopted by 't Hooft \cite{'thooftansatz} as we now demonstrate \cite{Manton}. The ansatz is that a self-dual gauge field can be constructed by use of a scalar function  $\Omega$ as  
\be 
\label{4.1} 
gA_{I} = \frac{1}{2}\sigma_{IJ}\partial_{J} \log \Omega, 
\ee 
where 
\be 
\label{4.2} 
\sigma_{IJ} = \frac{i}{2}\{\sigma_{I}\bar{\sigma}_{J} - (I \leftrightarrow J)\}; \ \ \ \sigma_{I} = (\vec{\tau}, iI), \ 
\bar{\sigma}_{I} = (\vec{\tau}, -iI).  
\ee
Note that the $\sigma_{IJ}$ is anti-self-dual in the sense $\sigma_{IJ} = - \frac{1}{2}\epsilon_{IJKL}\sigma_{KL}$, as is easily seen. Then the condition to be satisfied by a self-dual gauge field $F_{IJ} = \frac{1}{2}\epsilon_{IJKL}F_{KL}$ is 
\be 
\label{4.3} 
\sigma_{IJ}F_{IJ} = 0,  
\ee 
which, after some arithmetic manipulation, leads to a condition for the self-duality,   
\be 
\label{4.4} 
- 3 \{\partial_{I}^{2} \log \Omega + (\partial_{I} \log \Omega)^{2}\} I = 0 \ \to \ 
\partial_{I}^{2} \Omega = (\partial_{i}^{2} + \partial_{y}^{2})\Omega = 0. 
\ee 
Similarly, an anti-self-dual field can be constructed by replacing $\sigma_{IJ}$ in (\ref{4.1}) by $\bar{\sigma}_{IJ}$: 
\be 
\label{4.4'} 
gA_{I} = \frac{1}{2}\bar{\sigma}_{IJ}\partial_{J} \log \Omega \ \ \ (\bar{\sigma}_{IJ} = \frac{i}{2} \{ \bar{\sigma}_{I}\sigma_{J} - (I \leftrightarrow J) \}). 
\ee
 
In the case of the space-like instanton, since the field configuration preserves the combined symmetry of external and internal SO(4) rotation, it is expected to be generated by $\Omega$, which is the function of SO(4) invariant $\rho = \sqrt{x_{I}^{2}} = \sqrt{x_{i}^{2} + y^{2}}$. The relevant choice is 
\be 
\label{4.5} 
\Omega = 1 + \frac{\lambda^{2}}{\rho^{2}},  
\ee 
which clearly satisfies (\ref{4.4}): $\partial_{I}^{2} (\frac{1}{\rho^{2}}) = 0$. We adopt anti-self-dual configuration 
to see whether this method recovers the anti-instanton derived in (\ref{2.10a}) and (\ref{2.10b}).  Since $\partial_{J} \log \Omega 
= 2 \frac{x_{J}}{\rho^{2} + \lambda^{2}} - (\lambda = 0)$, (\ref{4.4'}) yields gauge fields, 
\be 
\label{4.6}
gA_{I} = \frac{\bar{\sigma}_{IJ}x_{J}}{\rho^{2} + \lambda^{2}} - (\lambda = 0). 
\ee
Actually, the term $(\lambda = 0)$, singular at the origin $\rho = 0$, turns out to be a pure gauge contribution, since it can be written as $iU \partial_{I} U^{\dagger}$ by use of $U$ in (\ref{2.10}), as is easily seen. So we now perform a gauge transformation due to $U$, in order to eliminate the $(\lambda = 0)$ term. The remaining term is also modified by this transformation and the resultant gauge field is written as 
\be 
\label{4.6'}
gA'_{I} = U^{\dagger} (\frac{\bar{\sigma}_{IJ}x_{J}}{\rho^{2} + \lambda^{2}}) U =  - \frac{\sigma_{IJ}x_{J}}{\rho^{2} + \lambda^{2}}. 
\ee 
Here we have used a relation $(\hat{x}_{K}\sigma_{K}) \bar{\sigma}_{IJ} (\hat{x}_{K'}\bar{\sigma}_{K'}) = - \sigma_{IJ}$.   
More explicitly, 
\bea 
&& g\vec{A}' = - \frac{y\vec{\tau} + \vec{\tau} \times \vec{x}}{\rho^{2} + \lambda^{2}},  
\label{4.6a}
\\ 
&& gA'_y =  \frac{\vec{\tau}\cdot \vec{x}}{\rho^{2} + \lambda^{2}},      
\label{4.6b}
\eea 
which are nothing but (\ref{2.10a}) and (\ref{2.10b}) obtained in the previous section.

In the case of the BPS monopole, which appears in ordinary 3D space after the naive dimensional reduction with only KK zero mode being kept, the gauge fields do not depend on the extra space coordinate $y$ and preserve only the combined symmetry of external and internal SO(3) rotation. Hence $\Omega$ should respect the symmetry and we adopt 
\be 
\label{4.7} 
\Omega = \frac{\sinh \tilde{r}}{\tilde{r}}e^{i\tilde{y}} \ \ (\tilde{r} = gvr, \ \tilde{y} = gvy).   
\ee 
It is easy to confirm $\partial_{I}^{2} \Omega = 0$. Actually the choice $\Omega = \frac{\cosh \tilde{r}}{\tilde{r}}e^{i\tilde{y}}$, e.g., also 
works as well. The reason to choose (\ref{4.7}) is that it is non-singular even at the origin $\tilde{r} = 0$.   

We again try an anti-self-dual gauge field, 
\be 
\label{4.8} 
gA_{I} = \frac{1}{2}\bar{\sigma}_{IJ}\partial_{J} \log \Omega.   
\ee 
Substituting (\ref{4.7}) in (\ref{4.8}), we obtain 
\bea 
&& A_{i} = - \frac{v}{2} \{ i\tau_{i} + (\coth \tilde{r} - \frac{1}{\tilde{r}}) \epsilon_{ijk} \hat{x}_{j}\tau_{k} \}
\label{4.9a} 
\\ 
&& A_y =  (\coth \tilde{r} - \frac{1}{\tilde{r}})v \hat{x}_{i} (\frac{\tau_{i}}{2}).     
\label{4.9b}
\eea 
The obtained gauge fields are $y$-independent, as they should be, though $\Omega$ itself depends on $y$. 

We, however, encounter a problem, i.e. that $\vec{A}$ contains an anti-hermitian part proportional to $i \tau_{i}$. 
Thus, we perform ``complexified" gauge transformation with a complex gauge parameter \cite{Manton2}, in order to eliminate this redundant term. 
The complexified gauge transformation is due to 
\be 
\label{4.10} 
U_{c} = e^{i\theta \tilde{x}_{j}\cdot \tau_{j}} \ \ \ (\tilde{x}_{i} = gv x_{i}), 
\ee 
where $\theta$ is a complex constant parameter and therefore $U_{c}^{\dagger} \neq U_{c}^{-1}$. Let us note that under the complexified gauge transformation 
\be 
\label{4.10a} 
A_{i} \ \to \ A'_{i} = U_{c}^{-1}A_{i}U_{c} + \frac{i}{g}U_{c}^{-1}\partial_{i}U_{c},  
\ee 
the field strength transforms as  
\be 
\label{4.10b} 
F_{ij} \ \to \ F'_{ij} = U_{c}^{-1}F_{ij}U_{c}. 
\ee 
Thus Tr $(F'_{ij}F'^{ij})$ = Tr $(U_{c}^{-1}F_{ij}F^{ij}U_{c})$ = Tr $(F_{ij}F^{ij})$ and the action is still invariant even under the complexified gauge transformation. This is why it is meaningful to consider the generalized gauge transformation. It may be worth mentioning that this generalized transformation makes sense since the theory contains only gauge fields. Once a complex matter field, such as $\psi$ belonging to the fundamental representation of SU(2), is introduced the theory is no longer invariant under the complexified  gauge transformation $\psi \to \ \psi' = U_{c} \psi$. Namely, $\bar{\psi'} \gamma^{i}D'_{i}\psi' = \bar{\psi}U_{c}^{\dagger} \gamma^{i} U_{c} D_{i}\psi \neq \bar{\psi} \gamma^{i}D_{i}\psi$, as $U_{c}^{\dagger} \neq U_{c}^{-1}$.

After the gauge transformation (\ref{4.10a}),  
\bea 
A'_{i} &=& - \frac{v}{2} \{ i[\tau_{i} - \sin (2\theta \tilde{r})\epsilon_{ijk}\hat{x}_{j}\tau_{k} - (1-\cos(2\theta \tilde{r}))(\delta_{ij} - \hat{x}_{i}\hat{x}_{j})\tau_{j}]  \nonumber \\ 
&+&  (\coth \tilde{r} - \frac{1}{\tilde{r}})[\cos (2\theta \tilde{r})\epsilon_{ijk}\hat{x}_{j}\tau_{k} + \sin (2\theta \tilde{r})(\delta_{ij} - \hat{x}_{i}\hat{x}_{j})\tau_{j}] \}  \nonumber \\ 
&+& v \{ - \theta \tau_{i} + [\theta - \frac{\sin (2\theta \tilde{r})}{2\tilde{r}} ] (\delta_{ij} - \hat{x}_{i}\hat{x}_{j})\tau_{j} + \frac{1 - \cos (2\theta \tilde{r})}{2\tilde{r}} \epsilon_{ijk}\hat{x}_{j}\tau_{k} \}. 
\label{4.11}
\eea 
Then, choosing a specific value $\theta = - \frac{i}{2}$, we can easily see that at the right hand side of (\ref{4.11}) all unnecessary anti-hermitian parts including the term $i\tau_{i}$ vanish (note that $\sin (2\theta \tilde{r}) = - i \sinh \tilde{r}, \ \cos (2\theta \tilde{r}) = \cosh \tilde{r}$ ) and we are left with hermitian gauge field, as we desire:   
\be 
\label{4.12} 
A'_{i} = (1 - \frac{\tilde{r}}{\sinh \tilde{r}}) \frac{1}{gr} \epsilon_{ijk}\hat{x}_{j}(\frac{\tau_{k}}{2}).  
\ee 
On the other hand, $A_{y}$ is clearly invariant under this gauge transformation, since $[ A_{y}, U_{c}] = 0$ and also $U_{c}$ is $y$-independent. The obtained (\ref{4.9b}) and (\ref{4.12}) are nothing but the results of (\ref{3.18}) and (\ref{3.19}) in the previous section, which were constructed in a different approach.

Although the relation between the space-like instanton and the BPS monopole becomes manifest through a unified description of anti-self-dual gauge field by use of the ansatz, as we have seen above, there still remains some essential difference between these two topological solitons. Namely, in the case of space-like instanton, SU(2) gauge symmetry is not broken spontaneously and the winding number characterizes the mapping from $S^{3}$ to the full gauge space SU(2), while in the  case of the BPS monopole, the SSB SU(2)$\to$U(1) plays a crucial role and the winding number characterizes the mapping from $S^{2}$ to the coset space SU(2)/U(1). 

It may be meaningful to study how these solitons behave at short range around the origin, since we naively expect that these two solitons show similar behaviors. This is because shorter distance corresponds to higher energy, while at higher energy we naively expect that the effect of the VEV causing the SSB may be safely ignored. 
At the limit of $\rho \to 0$ (also setting $y = 0$ as suggested by the naive dimensional reduction), the anti-self -dual field (\ref{2.10a}) and (\ref{2.10b}) for the space-like instanton behaves as 
\be 
\label{4.13} 
gA_{i} \simeq \frac{1}{\lambda^{2}} \epsilon_{aij}x_{j}\tau_{a}, \ \ \ gA_{y} \simeq  \frac{1}{\lambda^{2}} x_{a} \cdot\tau_{a}.    
\ee 
On the other hand, at the limit $r \to 0$, where $F(r) \to \frac{1}{3}\tilde{r}, \ G(r) \to \frac{1}{6}\tilde{r}^{2}$ (see 
(\ref{3.19})), the anti-self -dual field (\ref{3.18}) for the BPS monopole behaves as 
\be 
\label{4.14} 
gA_{i} \simeq \frac{1}{6}(gv)^{2} \epsilon_{aij}x_{j}(\frac{\tau_{a}}{2}), \ \ \ gA_{y} \simeq  \frac{1}{3} (gv)^{2} x_{a} \cdot (\frac{\tau_{a}}{2}).     
\ee 
So we realize that identifying $gv$ of the order of aforementioned $M_{X}$ with $\frac{1}{\lambda}$, these two solitons behave in a similar manner, though the complete identification is not possible because of the difference of factor 2, whose origin has not been understood.

\section{Summary and discussion} 

In the scenario of gauge Higgs unification (GHU) as a candidate of physics beyond the standard model, the Higgs field has an interpretation as a sort of Aharonov-Bohm phase, and the magnetic property together with the non-trivial topological nature of the extra dimension plays a crucial role in its foundation. 

In this paper we considered another magnetic property of this scenario. Namely, we considered 't Hooft-Polyakov monopole (TPM) as a topological soliton in the 5D GHU theory. The GHU provides a very suitable framework to incorporate the TPM, since the adjoint scalar, usually introduced as a matter field in 4D gauge theory, is built in in the GHU as the extra-space component of higher dimensional gauge field. In other words, the presence of the TPM is in some sense inevitable in this scenario, especially when the gauge group is simple, such as SU(3) adopted in the case of the minimal GHU electro-weak unified model \cite{Kubo}. 

In the process of the analysis, we realized that the condition to be satisfied by the BPS state of the TPM, the BPS monopole, is equivalent to a (anti-)self-dual condition, subject to the naive dimensional reduction, for the higher dimensional gauge field. This observation, in turn, strongly suggests the presence of an instanton-like topological soliton living in the 4D space  including the extra dimension, instead of the 4D space-time in the case of the ordinary instanton, say ``space-like instanton" with finite energy (therefore a mass), instead of finite action in the case of the ordinary instanton.  
We constructed the field configuration for the space-like instanton as an anti-self-dual gauge field and calculated its mass, which is of the order of $M_{c}/\alpha$ with $M_{c} = 1/R \ (R: {\rm the \ size \ of \ the \ extra \ dimension})$ being the compactification mass scale.  

Next we discussed TPM, especially the BPS monopole, as an anti-self-dual gauge field obtained under the naive dimensional reduction, with only the KK zero modes being kept. We started from the attempt to construct a hedgehog-type solution with winding number 1 as what is obtained by a local gauge transformation from the trivial vacuum state, in order to get a finite energy solution. 
The mass of the BPS monopole was also estimated to be of the order of $M_{X}/\alpha$, with $M_{X}$ being the mass scale of the spontaneous symmetry breaking (SSB) due to the VEV, which is supposed to be much larger than the weak scale $M_{W}$. For instance, in the realistic SU(3) GHU model, $M_{X}$ characterizes the SSB, SU(3)$\to$SU(2)$_{L} \times $U(1)$_Y$. 

Finally we further discussed the relationship between the space-like instanton and the BPS monopole. These two topological solitons should have close relation with each another, since both gauge field configurations are interpreted as anti-self-dual fields, and also homotopy classes behind these topological solitons are identical. 

We argued that the relationship becomes manifest through an unified description of anti-self-dual gauge field by use of the ansatz
 adopted by 't Hooft \cite{'thooftansatz}, \cite{Manton}. The ansatz is described in terms of one scalar function $\Omega$. We demonstrated that if we adopt the $\Omega$, which is invariant under the SO(4) rotation of the 4D space (including the extra dimension), we naturally obtain the space-like instanton solution, while if we adopt the $\Omega$, which is invariant under the SO(3) rotation of the ordinary 3D space, we naturally obtain the BPS monopole solution. 
We also discussed briefly that at shorter distances, where the effect of the vacuum expectation value is expected to be relatively unimportant, the behaviors of gauge field configurations of these two-types of soliton are similar.   

Although the relationship has been established through the unified description by use of the ansatz, it is still not completely clear how the field configuration of the space-like instanton is smoothly modified into that of the BPS monopole under the naive dimensional reduction. One may wonder what happens if we pick up the KK zero mode of anti-self-duality condition, $F_{IJ} = - \frac{1}{2}\epsilon_{IJKL}F_{KL}$, satisfied by the field configuration of the space-like instanton. The point here is that in the KK zero mode of the field strength $F_{IJ} = \partial_{I} A_{J} - \partial_{J}A_{I} - ig [A_{I}, A_{J}]$, KK non-zero modes also participate through the commutator as $\sum_{n}[A^{(n)}_{I}, A^{(-n)}_{J}]$ ($n$: KK modes),  while in the case of the BPS monopole, the anti-self-duality condition should be satisfied only by the $y$-independent gauge field, i.e. only by the KK zero mode. Thus, to establish the relationship between the KK zero mode of the space-like instanton and the BPS monopole solution is not straightforward. We leave this problem together with other remaining problems for future investigations.

\subsection*{Acknowledgments}

This work was supported in part by Japan Society for the Promotion of Science, Grants-in-Aid for Scientific Research, No.~15K05062, No.~16H00872.


\providecommand{\href}[2]{#2}\begingroup\raggedright\endgroup

\end{document}